\documentclass[preprint]{aastex}
\input{psfig.tex}  

\newcommand{\gsim}{\;\lower.6ex\hbox{$\sim$}\kern-7.75pt\raise.65ex\hbox{$>$}\;}
\newcommand{\lsim}{\;\lower.6ex\hbox{$\sim$}\kern-7.75pt\raise.65ex\hbox{$<$}\;}

\begin{document}

\title{CU COMAE: A NEW FIELD DOUBLE-MODE RR LYRAE, THE MOST METAL POOR 
DISCOVERED TO DATE
\footnote{Based on data obtained with the 1.52 m telescope of the Bologna 
Observatory in Loiano, the Southwestern University 40 cm telescope, 
and the 2.7 m telescope of the 
McDonald Observatory} }

\author{G. Clementini\altaffilmark{2}, S. Di Tomaso\altaffilmark{3}, 
 L. Di Fabrizio\altaffilmark{2}, A. Bragaglia\altaffilmark{2}, 
 R. Merighi\altaffilmark{2}, M. Tosi\altaffilmark{2}, 
 E. Carretta\altaffilmark{4}, R.G. Gratton\altaffilmark{4}, 
 I.I. Ivans\altaffilmark{5}, A. Kinard\altaffilmark{6}, 
M. Marconi\altaffilmark{7}
H.A. Smith\altaffilmark{8}, R. Wilhelm\altaffilmark{6}, T. 
Woodruff\altaffilmark{6}, C. Sneden\altaffilmark{5}}
\altaffiltext{2}{Osservatorio Astronomico di Bologna, Via Ranzani 1, 
  I-40127 Bologna, Italy}
\altaffiltext{3}{Dipartimento di Astronomia, Universit\`a di Bologna, Via 
Ranzani 1, 
 I-40127 Bologna, Italy}
\altaffiltext{4}{Osservatorio Astronomico di Padova, Vicolo dell'Osservatorio 
5, I-35122 Padova, Italy}
\altaffiltext{5}{Department of Astronomy, University of Texas at Austin, 
Austin, TX
78712-1083, USA}
\altaffiltext{6}{Department of Physics, Southwestern University, 
Georgetown, TX 78627, USA}
\altaffiltext{7}{Osservatorio Astronomico di Capodimonte, Via Moiariello 16, 
I-80131 Napoli, Italy}
\altaffiltext{8}{Department of Physics \& Astronomy, Michigan State University, 
East Lansing, Michigan 48824, USA}
\email{gisella@bo.astro.it, 
 s\_ditomaso@astbo4.bo.astro.it, s\_difabrizio@astbo4.bo.astro.it, 
 angela@bo.astro.it, merighi@bo.astro.it, tosi@bo.astro.it, 
 carretta@pd.astro.it, gratton@pd.astro.it, iivans@astro.as.utexas.edu, 
 kinard@southwestern.edu, marcella@cerere.na.astro.it, 
 smith@saucer.pa.msu.edu, wilhelm@nokomis.as.utexas.edu, 
 woodruff@southwestern.edu, chris@verdi.as.utexas.edu}


\begin{abstract}
We report the discovery of a new double-mode RR Lyrae variable (RRd) 
in the field of our Galaxy: CU Comae. CU Comae is the sixth such RRd 
identified to date and is the most metal-poor RRd ever detected. 
Based on BVI CCD photometry spanning eleven years of
observations, we find that CU Comae has periods P$_0$=0.$^d$5441641($\pm 
0.0000049$) and P$_1$=0.$^d$4057605($\pm 0.0000018$).
The amplitude of the primary (first-overtone) period of CU Comae
is about twice the amplitude of the secondary (fundamental) period. 
The combination of the fundamental period of pulsation P$_0$
and the period ratio of 
P$_1$/P$_0$=0.7457 places the variable on the metal-poor side of the Petersen 
diagram, in the
region occupied by M68 and M15 RRd's. A mass of 0.83 M$_\odot$ is estimated for
CU Comae using an updated theoretical calibration of the Petersen diagram.
High resolution spectroscopy (R=30,000) covering one full pulsation cycle 
of CU Comae was obtained with the 2.7 m telescope of the Mc Donald Observatory,
and has been used to build up the radial velocity curve of the variable.
Abundance analysis done on the four spectra taken near 
minimum light ($0.54 < \Phi < 0.71$) confirms the metal poor nature of CU 
Comae, for which 
we derive [Fe/H]=$-2.38 \pm 0.20$. This value places this new RRd  
at the extreme metal-poor edge of the metallicity distribution of the RR Lyrae 
variables in our Galaxy. 
\end{abstract}


\newpage

\section{Introduction}

Double-mode RR Lyrae variable stars (RRd) play a fundamental r\^ole in
astrophysics because they can be used to estimate the mass and the
mass-metallicity relation of horizontal branch stars, and because they
can provide information on the direction and rate of evolution across the
horizontal branch. RRd's are variables which pulsate both in the fundamental 
and first overtone radial pulsation modes. The  evolutionary interpretation of 
the double-mode
phenomenon suggests that these stars are changing their pulsation mode
while evolving across the RR Lyrae instability strip. 
The mixing of the two pulsation modes produces cycle-to-cycle amplitude 
changes and a large
scatter in the observed light curves, much larger than accounted for by
observational errors alone. 
First discovered in 1977 (AQ Leo, Jerzykiewicz \&
Wenzel 1977), about 40 RRd's have been detected so far in a number of  
globular clusters of our Galaxy (e.g. M15: [Fe/H] = --2.12, M68:
[Fe/H] = --1.99, IC4499: [Fe/H] = --1.26 on the Carretta \& Gratton 1997, CG97,
metallicity scale). Less than a hundred are hosted in a number of dwarf 
spheroidal
(Draco: [Fe/H] = --1.83, Sculptor: [Fe/H] = --1.58, from Mateo 1998, transformed
to the CG97 metallicity scale) and irregular galaxies of the
Local Group; in particular 73 double-mode RR Lyraes have recently been
detected by the MACHO experiment (Alcock et al. 1997) in the Large Magellanic
Cloud (LMC). However, double-mode RR Lyraes  seem to be a rather rare event in
the field of our Galaxy where only 5 RRd's have previously been detected 
(Jerzykiewicz \& Wenzel 1977; Clement, Kinman \& Suntzeff 1991; Garcia-Melendo
\& Clement 1997; Moskalik 1999). 
All of the double-mode RR Lyraes
have been so far generally found in metal poor stellar systems ([Fe/H] 
$\lsim -1.3$
dex, CG97 scale, or --1.5 dex, Zinn \& West 1984 scale).
The metal abundance estimated by Clement et al. (1991) for three of the 5
Galactic RRds known so far (--1.70, --1.72, --1.60 for AQ Leo, RR VIII-58 and
RR VIII-10 respectively, on the CG97 metallicity scale,
corrected from $-1.90$, $-1.91$ and $-1.81$, on Zinn \& West 1984 scale)
confirms this finding. However, since the metallicity distribution of the RR
Lyrae variables in the Galaxy extends to metal abundances both much lower and
higher than in globular clusters, it is very important to identify RRd's
among the field RR Lyrae population in order to test the mass-metallicity
relation on a much wider metallicity interval. Applying the $\Delta$S method 
(Preston 1959) to  
low resolution spectroscopic observations of 
RRd's in the bar of
the LMC, Bragaglia et al. (2000) 
inferred metallicities in the range of
$-1.09$ to $-1.78$ dex (or to $-2.24$ including one star with lower weight),
and confirm that the LMC field RRd's follow the mass-metallicity
relation defined by the Galactic cluster double-mode pulsators.
However, with so few RRd stars identified in the field of our 
Galaxy, it is not possible to assess whether Galactic field RRd's 
actually obey the same mass-metallicity relation defined by the 
globular cluster RRd's. The discovery of any new Galactic field RRd helps set 
important constraints on the mass-metallicity relation; the discovery of this 
particular extremely metal-poor RRd also sets 
a new boundary condition.

	In this paper, we present results based on new photometric data for
	the RR Lyrae variable CU Comae (CU Com) taken from 1995 to 1999, 
	combined with published photometry of this object taken from
	1989 to 1994 (Clementini et al. 1995b), together with high resolution 
	spectroscopy obtained in 1999, that covers an entire pulsation cycle.  
	Section 2 presents the observations and data sets.  In section 3, we 
	give our results from the analysis of the entire photometric 
	data-set of CU Com (467 V, 172 B, and 167 I data points), which 
	spans 11 years of observations.  In Section 4 we report the results 
	of our spectroscopic analyses.  We have both derived a	radial 
	velocity curve of the full pulsation cycle (Section 4.1) and 
	performed an elemental abundance analysis of the spectra of CU Com 
	taken near minimum light (Section 4.2).  In section 5, we estimate 
	the mass of CU Com using a new theoretical calibration of the Petersen 
	diagram (Petersen 1973), summarize the main derived quantities of CU Com, and 
	discuss the impact of this new discovery on both the mass 
	metallicity relation, and on the evolutionary interpretation of the 
	double-mode pulsation for RR Lyrae variables. 

\section{Observations and reductions}

In the fourth edition of the General Catalogue of Variable Stars (Kholopov et
al. 1985, GCVS4) CU Com ($\alpha_{2000}=12^h 24^m 47^s$, 
$\delta_{2000}=22^o 24' 29''$) is classified as an {\it ab} type RR Lyrae
with P=0.$^d$416091 and amplitude of the photographic light variation of 0.5 
mag.
Clementini et al. (1995b) presented observations of CU Com taken during
the years 1989-1994. Their V light curve (121 data points) has a sinusoidal
shape (see their fig. 3b) reminiscent of a {\it c} type pulsator; however,
the region around maximum light is split into two separate branches about
0.15 mag apart. They also derived a shorter period of P=0.$^d$405749 and a 
larger
(V) amplitude of 0.58 mag compared to the GCVS4 values. The irregular nature
of the CU Com light curve is confirmed by the much less 
sampled photometry of Schmidt, Chab \& Reiswig (1995, 26 data points, 
see their fig. 2).
However, neither the Clementini et al. (1995b), nor Schmidt et al. (1995), 
photometries were sufficient to address the issue of whether CU Com 
was exhibiting double-mode or non-radial mode pulsation (Olech et al. 1999), 
or whether it
might be affected by the Blazhko effect (Blazhko 1907). 
A new observing campaign was conducted on CU Com from 1995 to 1999, 
collecting data on several
consecutive nights of each run and in two-three runs about one month apart,
to test both the possibility of double-mode or non-radial mode pulsation 
(which both are known to occur on cycle-to-cycle timescales), 
and of
the Blazhko effect (whose periodic modulation of the lightcurve 
typically has timescales of tens of days).
Moreover, during the 1999 campaign, coordinated observations at the 1.52 m
telescope of the Bologna Observatory in Loiano, 
at the 60 cm of the Michigan State University, and at the 
40 cm  of the Southwestern University were organized, to obtain continuous 
photometry with time 
coverage longer than 12-16 hours.
High resolution spectroscopy was also secured in order to check whether CU Com
 might be the component of a spectroscopic binary system, to obtain its radial
velocity curve, and to perform an abundance analysis of the variable.
Uneven weather conditions at the various sites prevented us from getting 
long-span photometric observing nights, however we succeeded to obtain 
simultaneous photometry and spectroscopy with the 40 cm Southwestern 
University and the 2.7 m McDonald telescopes, on the night of 1999, February 
12.
\subsection{The photometry}

The new photometric observations of CU Com were obtained in 21 nights from 
 March 1995 to April 1999. 
They consist essentially of BVI CCD
observations in the Johnson-Cousins system obtained with the Loiano 1.52 m 
telescope operated by the Bologna Observatory, and 
are complemented by
17 V and 8 B frames obtained with the Southwestern University 40 cm.
The journal of the new photometric observations is given in Table~1.
Observations at the 1.52 m telescope were obtained with two different 
instrumental set-ups, (i) with an RCA CCD for direct imaging having 
a 4.3 $\times$ 2.7 
arcmin$^2$ field of view and a 0.5 arcsec/pixel scale, and (ii) with BFOSC
(Bologna Faint Object Spectrograph \&  Camera) mounting a Thomson 1k$\times$1k
CCD with 0.5 arcsec/pixel scale giving a field of view of 9.6 $\times$ 9.6 
arcmin$^2$. 
A filter wheel for the  Johnson-Cousins photometric system was used with 
both set-ups.
A 2.3 $\times$ 2.8 
arcmin$^2$ CCD image of
the CU Com field is shown in Figure~1.  
Four stars, beside CU Com, are marked on the image. These are non 
variable objects which were used as reference stars. The light curve of CU 
Com is derived in terms of differential 
magnitude with respect to these comparison stars (whose constancy has been 
accurately checked), 
hence no concern arises when observations were performed in non strictly
photometric conditions. 
Observations at the 40 cm Southwestern University telescope were obtained
using standard Johnson B and V filters and a Pictor 416XT CCD camera coupled to
a Celestron f/6.3 focal reducer.  This configuration gives a scale of 0.8 
arcsec/pixel resulting in a 7.4 $\times$ 10.1 arcmin$^2$ field of view.
Data were pre-reduced 
and instrumental magnitudes of the variable and its comparison stars were 
derived by direct photon counting 
using standard routines for aperture 
photometry in IRAF\footnote{IRAF is distributed by the National Optical 
Astronomy Observatories, which is operated by the Association of Universities 
for Research in Astronomy, Inc., under cooperative agreement with the 
National Science Foundation.}. 
The photometric data were tied to the standard photometric system
through observation of 35 standard stars selected from 
Landolt (1983, 1992). 
Calibrated magnitudes of the 4 comparison stars are given in 
Table~2. Comparison stars 
are identified in column 
2 of the table by their numbers on the 1.2 version of the {\it Hubble Space 
Telescope} Guide Star
Catalogue (GSC1.2).
Within the quoted photometric uncertainty the V magnitude of star C1 given in 
Table~2 agrees with the value published by Clementini et al. (1995b), however 
the present value, being the average of several 
measurements obtained in four independent calibration nights, supersedes the 
1995 estimate.
The B, V, and I magnitudes of CU Com relative to the primary comparison star 
C1 
are listed in Table~3 along with the Heliocentric Julian date at 
mid-exposure. Only data corresponding to the first night of observations 
of the 1995-1999 span are listed in the table. According to the errors quoted 
in Table~2 we estimate that 
the photometric accuracy of each individual data point is 
$\pm 0.03$ mag in V and B, and $\pm 0.04$ mag in I.
For sake of homogeneity, instrumental magnitudes of CU Com and of its 
comparison stars in the 1989-1994 data 
published by Clementini et al. (1995b), were all re-measured in order to 
reduce any systematic effect which might arise in the analysis of the 
photometric data from inhomogeneities in the magnitude measuring procedure.
The full photometric data set including the re-measured Clementini et al 
(1995b) photometry is published in the electronic edition of the Journal.

\subsection{The spectroscopy}
CU Com is the first double-mode RR Lyrae for which high resolution
spectroscopy is available. Fifteen spectra evenly covering 
the full pulsation cycle of CU Com  were obtained
with the cross-dispersed echelle spectrometer at the 
coud\'e f/32.5 focus of the Harlan J. Smith 2.7 m 
telescope at the McDonald
Observatory in West Texas during the nights of 1999, February 12 and 13.
Seeing conditions during the nights varied from 1.5$^{\prime \prime}$ to 
1.7$^{\prime \prime}$.
Exposures of a Thorium-Argon lamp were taken at beginning, middle and end of 
each night, in order to secure the absolute
wavelength calibration. Observations of the radial velocity standard
star  HD 58923 (RV=+17.8 km s$^{-1}$, Wilson 1953) were also obtained in 
order to 
use them 
during the cross-correlation measure of the radial velocities 
of CU Com and to set
the radial velocity zero point.

	The ``2d-coud\'e'' spectrograph (Tull et al. 1995) was operated 
	using the E2 echelle grating (52.6759 grooves mm$^{-1}$ with a blaze 
	angle of 65.293) which yields a two-pixel resolving power at the 
	CCD of R=$\lambda/\Delta \lambda$=63 000.  Our wavelength coverage 
	extended from 3670\AA\ to 9900\AA\ (with inter-order gaps in 
	wavelength coverage redward of 5600\AA) with the data from 
	about 4400 to 5300 \AA, and from 3924 to 8465 \AA\ 
 used in the radial velocity and metallicity 
	determinations, respectively.  

	Given the faintness of our target ($<$V$> \sim 13.3$ and V$_{min}
	\sim$ 13.6 mag) and the severe constraints on the exposure length 
	(exposures of CU Com could not exceed 20-30 min in order to avoid 
	phase blurring on the pulsation cycle of this short period variable 
	star), we used a 1010 $\mu$m slit (which projects to 2 arcsec on
	the sky) and performed a 2$\times$2 binning during the read-out of 
	the CCD in order to increase the S/N ratio.  The FWHM measured 
	from the ThAr comparison lamp lines is $\sim$ 0.2\AA.  The
	data were taken on a 2048$\times$2048 thinned, grade 1 Textronix 
	CCD.

	Data were reduced using standard IRAF processing tools to correct 
	and calibrate for various detector effects.  
   The overscan region of each image was used to determine the bias level
to subtract from each frame. 
 We then trimmed the overscan 
	region from the 
	individual images.  Identification of the locations of ``bad'' 
	pixels was made by means of long exposure dark frames as well as 
	flat field frames.  Bad pixel values were replaced using 
	interpolated values from neighbouring pixels.  We then divided the 
	stellar and comparison lamp spectra by a scaled ``flat field'' 
	image (comprised of 20 spectra taken with a quartz lamp through a 
	blue-pass filter) in order to correct for the relative differences 
	in pixel response of the detector.  The minimum order separation 
	was 10 arcsec, leaving sufficient interorder background free of 
	contamination from starlight in order to adequately remove the
	(small) scattered light contribution.  Cosmic ray excision was 
	performed using {\it lineclean}.  

	We extracted the flux in the individual orders of the two-dimensional 
	data to obtain one-dimensional summed spectra of the stellar and 
	comparison lamp spectra.  Using the Th-Ar comparison lamp emission 
	spectra, we determined and applied the dispersion relation to the 
	stellar spectra, interpolating in time over shifts in the comparison 
	line positions on the chip through the course of a night.  We 
	verified that the resulting wavelength solutions gave correct 
	positions of known atmospheric emission and absorption features on 
	each spectrum.  Geocentric and heliocentric Julian dates were 
	computed and the information implanted in the file headers.  The 
	information was then used to compute the corrections to the observed 
	radial velocity due to the diurnal, monthly, and yearly motions.

\section{Analysis of the photometric data}
Using the differential photometry of CU Com with respect 
to the reference star C1 and the full 1989-1999 data-set 
(the present new photometry together with Clementini et al, 1995b, one) 
we performed a 
period search with GRATIS (GRaphycal Analyzer
TIme Series), a code being developed at the Bologna Observatory 
which employes two
different algorithms:    (a) a Lomb periodogram  (Lomb 1976, Scargle 1982)
and (b) a best-fit of the data with a truncated Fourier series (Barning
1962).
  The adopted period search procedure was to perform the Lomb analysis on    a
wide period interval first, and then to use the Fourier algorithm to    refine
the period definition and  find the best fitting model.
The period search employed each of the complete B, V and I data-sets.
Figure~2 shows the periodograms of CU Com V, B, and I data
respectively, obtained using the Lomb algorithm to identify the most 
probable frequency of the data on a wide period interval of 0.2-0.7 day.
The highest peaks at $\omega$=2.46 (P$\sim$0.406 d) correspond to the primary
periodicity of the data; the two lower peaks at $\omega$=1.46 and
$\omega$=3.46 respectively, are aliases of the primary periodicity, while the 
peaks at $\omega$=1.84 are the signature of a second
periodicity present in the data.

We then reduced the interval around the primary periodicity using 
the Fourier algorithm to find the best fit. The  period obtained using a
three harmonics best-fitting Fourier series on the V data, and a two harmonics
series on the B and I data is P=0.$^d$405759$\pm$0.000001. 
Shown in Figure~3, are the V, B, and I  light
curves of CU Com obtained phasing the data according to this periodicity,
together with the best fitting models. Filled circles mark data obtained 
with the 40 cm Southwestern University telescope.

The {\it r.m.s.} deviation 
from the best-fitting model is $\pm 0.083$ mag in V, $\pm 0.109$ mag in B 
and $\pm 0.056$ mag in I. These residuals are much larger than expected from 
observational errors alone ($\pm 0.03$ mag in V and B, and  $\pm 0.04$ mag 
in I) and provide us with clues that a second periodicity may be present in the 
data.
The amplitude of the light variation ranges from 0.63 to 0.30 mag in V, 
from 0.82 to 0.45 in B, and  from 0.41 to 0.20 in I. 
Data corresponding to consecutive nights show that the amplitude variation 
takes place  on a 
cycle-to-cycle timescale (see Di Tomaso, 2000 for details). 
This occurrence definitely rules out the 
Blazhko effect as explanation of the irregular behaviour of CU Com.
Since 
the residuals and amplitude of the light
variation of CU Com cannot be described by a single periodicity, and the 
variation occurs on cycle-to-cycle basis, 
we proceeded looking for a second periodicity in the data on timescales
of the order of one day.
Data were prewhitened using the best fitting models in Figure~3 
and a new period search was performed on the residuals with respect to the 
models. 
Figure~4 shows 
the periodograms of the residuals of the V, B and I data, respectively, 
 obtained from the Lomb algorithm on a wide period interval of 0.1-0.9 d.
Strong peaks are present at $\omega$=1.84 (P$\sim$0.54 d) in all the 
three periodograms, which clearly mark the 
second periodicity present in the data of CU Com. 
We further refined the secondary period by restricting the 
period search interval and using a 3 harmonics best-fitting Fourier series in V
and a 2 harmonics model in B and I, and then performed an iterative 
refinement of both primary and
secondary periodicities. At the end of the trial procedure 
(conducted on the V, B, and I data, independently) the final adopted 
periodicities of CU Com are: 
P$_1$=0.$^d$4057605$\pm$ 0.0000018 and P$_0$=0.$^d$5441641$\pm$ 0.0000049.
We also conducted another, independent period search on the CU Com 
V data-set using the Date Compensated Discrete Fourier Transform program 
(DCDFT, Ferraz-Mello 1981) and the
CLEANEST routine (Foster 1995). Its results (P$_1$=0.$^d$405760$\pm$ 
0.000001 and P$_0$=0.$^d$544166$\pm$ 0.000003 with the DCDFT; 
P$_1$=0.$^d$405761$\pm$ 0.000001 and P$_0$=0.$^d$544164$\pm$ 0.000003 
with CLEANEST) fully
confirm the periodicities found by the GRATIS algorithms.

The ratio of the first to second period of CU Com 
is P$_1$/P$_0$=0.745658 $\pm$ 0.000007 which is
a typical value for double-mode RR Lyrae stars.
The top panels of Figure~5, 6 and 7 show the V, B and I light 
curves of CU Com 
phased according to the primary (first-overtone) period of pulsation 
P$_1$=0.$^d$4057605 and the epoch of maximum light E=2450142.$^d$5860390. 
The central panels give the light curves of the primary 
period after prewhitening of the secondary (fundamental) period 
P$_0$=0.$^d$5441641, and the bottom panels show the light curves of the 
secondary period after prewhitening of the primary periodicity.
The $r.m.s.$ deviations of the best fitting models of the prewhitened data 
(central panels of Figure~5, 6 and 
7) are about halved with respect to the 
original data (0.049 mag in 
V, 0.069 mag in B and 0.034 mag in I).
The amplitudes of the primary 
variation are 0.43 mag, 0.55 mag and 0.27 mag in V, B and I respectively, and 
the corresponding amplitudes of the secondary (fundamental) variation are 
0.22, 0.25 and 0.14 mag.
Since the B and V residuals are still slightly too large when compared to the
quoted photometric errors, we investigated whether a third periodicity might be
present in the 
data. Indeed, several low peaks 
around
P$\sim$0.$^d$233 were found, both with 
GRATIS and CLEANEST, when searching for this additional periodicity (P$_2$) 
the V, B and I data, 
independently.
The highest peaks of each individual  
data set give periods 
in the range 
0.$^d$232439--0.$^d$232587 ($\omega$=4.2995--4.3022), 
and the corresponding primary and secondary 
periodicities we found are in the ranges P$_1$=0.$^d$405757--0.$^d$405762 and 
P$_0$=0.$^d$544164--0.$^d$544173, respectively. While inclusion of these 
third periodicities, (after prewhitening each individual data set 
according to its 
P$_0$ and P$_2$ periodicities) marginally reduces the 
{\it r.m.s.} deviations of the best fitting models (from 0.049 to 0.040 mag, 
from 0.069 to 0.059, and from 0.034 to 0.030 in V, B and I, respectively),
residuals are almost identical if data are prewhitened according to 
second and third periodicities which are the average of the P$_0$ and P$_2$
values found from each of the three data set independently (0.047 mag in V, 
0.065 mag in B, and 
0.034 mag in I).
On the other hand,
these "large" residuals might partially be caused by the lower accuracy of some
of the photometric data (particularly data obtained in full moon nights with 
very 
poor seeing conditions). In order to investigate this possibility further 
we restricted the analysis to the data where the difference between comparison
stars (C1 and C4, in particular) remained constant within $\pm$0.03 mag, which 
is the typical photometric error both in V and B. The data-set was thus
restricted to 378, 116, and 134
 data-points in 
V, B, and I, respectively. While periodicities found with this data subset 
fully
confirm the results obtained with the complete data-set, the residuals are 
indeed 
slightly reduced (0.043 in V, 0.067 B, and 0.031 in I, respectively).
A search for a third 
periodicity was conducted also on this small data-set. Again several low peaks 
around
0.$^d$233 were found. 
Inclusion of a third period (which is the average of the individual P$_2$ 
values found from the V, B and I data, independently) 
 in the analysis
of the data reduces the residuals to the fit to 0.38 mag in V, 0.059 mag in B 
and 0.025 mag in I.
However, the patterns of the peaks in the 0.$^d$233 area are somewhat confused.
Overall, this third periodicity is not very strongly supported 
by the data.

\section{Analysis of the spectroscopic data}

\subsection{The radial velocity curve}
Radial velocities were measured from the reduced wavelength calibrated spectra
of CU Com using a cross correlation technique ({\it fxcor} in IRAF). Twelve
orders containing weak metal lines (the best suited to measure radial
velocities) in the spectra of CU Com were cross-correlated against the same
orders in the spectra of the radial velocity standard star HD 58923. Table~4
lists  the derived heliocentric radial velocities of CU Com and corresponding
errors, along with the heliocentric phases of the spectra, computed according
to the final adopted primary period of pulsation and the epoch of CU Com
(P$_1$=0.$^d$4057605 and E=2450142.5860390). Neither splitting of lines nor
double  cross correlation peaks were observed, which would otherwise provide a
hint of a spectroscopic binary system. Errors in the measured radial
velocities of CU Com (see last column of Table~4) are larger than found for 2
other RR Lyraes observed in the same observing run (CM Leo and BS Com, an RRc
and an RRab, respectively).
These slightly large uncertainties (typical errors of 3 instead of 
2 km s$^{-1}$, 
in spite of CU Com being about half magnitude brighter than CM Leo)
can be attributed to the extremely low metal abundance of the star 
(see Section 4.2).
The bottom panel of Figure~8 shows the radial velocity curve 
of CU Com obtained phasing the data according to its primary period 
of pulsation, while the top panel shows the simultaneous 
V light curve obtained with the 40 cm Southwestern University telescope.
The shape and amplitude of the radial velocity curve (A$_{\rm RV}$= 32.73 
km s$^{-1}$) 
are typical of a {\it c}-type pulsator. 
The systemic velocity ($\gamma$) of CU Com was calculated by integration 
of the radial velocity curve on the full pulsation cycle and corresponds to 
$-$54.13 km s$^{-1}$.

\subsection{The metallicity of CU Com}

Four spectra of CU Com taken at/near the minimum light (0.54$ < \Phi <$ 0.71) 
were used to
measure the metal abundance of the variable from the following line analysis
procedure.

First, we derived the effective temperature T$_{\rm eff}$\ from the dereddened
B$-$V and V$-$I colors of CU Com at minimum light. A reddening value of
E(B$-$V)=0.023 mag was estimated for CU Com from Schlegel, Finkbeiner \& Davis 
(1998) reddening maps. 
Table~5 lists the dereddened B$-$V
and V$-$I photometric colors corresponding to the four exposures of CU Com
close to minimum light, derived from the B and V 
light curves simultaneous to the spectroscopic data  and from the total I  
light curve (since no simultaneous I photometry is available), 
along with the corresponding temperatures 
as estimated using the color-temperature calibration and procedure of
Clementini et al. (1995a). 
The average {\it photometric} temperature we derive 
is T$_{\rm eff}=6286$~ $\pm$ 112~K where 
according to Clementini et al (1995a; see their Section 3.4.2)
the mean temperature derived from individual 
(B$-$V)$_0$'s was lowered by 49 K, and that from the (V$-$I)$_0$'s 
was lowered by 97 K, before averaging them together, to tie them to 
the V$-$K-temperature calibration for RR Lyrae stars, since 
this calibration is not affected by discrepancy between synthetic (Kurucs 
1993) and observed colors (see Figure 3 of Clementini et al, 1995a) and it is 
less metallicity dependent (see Fernley, 1989, and reference therein).
Within the errors, there is no difference between the 
mean temperature estimated using colors at minimum light derived 
from the total B, V, and  I light curves and the 
temperature obtained from the simultaneous photometry.

As an alternative approach, temperatures were also estimated by fitting the
profile of the H$\beta$ line (the H$\alpha$\ profile 
is usually used in this technique but 
was not observed with the adopted wavelength set up) in the four spectra. 
We devised the following effective technique to remove the 
characteristic wavelength response induced by the 
echelle  
blaze function, a pseudo-flat fielding was obtained by dividing the spectrum
of the order including H$\beta$\ by the average of the two adjacent orders
(this division was made using the pixel values, ignoring the wavelength
calibration); before this division was made, cosmic rays and strong absorption
lines (other than H$\beta$) were excised, and the spectra of the adjacent
orders were gaussian-smoothed with a FWHM of
1~\AA. This procedure works quite well because the instrumental response has no
strong wavelength dependence other than the echelle blaze function. The 
continuum was traced far from the center of 
 H$\beta$\ and normalized to a value of unity, and 
the H$\beta$\ feature and surrounding regions were then 
compared with
line profiles obtained following the same precepts described in Castelli,
Gratton \& Kurucz (1997).
Figure~9 shows the result of this comparison obtained using the
Kurucz (1993) model atmospheres, with the overshooting option switched on 
(these
models are used for consistency with the analysis of Clementini et al. 1995a;
temperatures obtained with the Kurucz model atmospheres without overshooting
are lower by about 250~K). The average {\it spectroscopic} temperature we 
obtained for the four
spectra taken close to minimum light is 6400$\pm$150~K. Within their quoted 
uncertainties photometric and spectroscopic temperatures agree well. 
	The temperature from H$\beta$\ is about 100 K higher than the 
	photometric one.  Adopting the photometric temperature in our 
	analysis permits us to stay on the same scale of Clementini et al. 
	(1995a), so that the new abundances can be directly compared with 
	those and avoids any lingering doubts about the effectiveness 
	with which we determined the continuum region near H$\beta$. 
Furthermore, the photometric temperature is less sensitive to the adopted 
set of model atmospheres.
We then summed up the 4 individual spectra (after excision of the cosmic rays
and shift to zero radial velocity). The coadded spectrum was convolved with a
Gaussian having a FWHM of 0.3~\AA\ to enhance the S/N and allows us to 
measure
faint, unsaturated lines with confidence. Finally, we measured equivalent 
widths of 15 Fe~I
and 8 Fe~II lines on the coadded spectrum.
Table~6 gives the linelist, the assumed $\log gf$'s and the corresponding 
EW's we measured. Only the first ten lines of the table are shown.
The full table is  
available in electronic form upon request to the first author.
A full discussion of the reliability of these adopted atomic parameters
to determine 
abundances of RR Lyrae variable stars can be found in Section 4.1.1 of
 Clementini et al. (1995a). 

We obtained average abundances of
[Fe/H]=$-2.35\pm$0.026 (with $\sigma$=0.10 dex) from Fe~I lines, 
and [Fe/H]=$-2.40\pm$0.028 (with $\sigma$=0.08 dex) from
Fe~II lines, using a stellar atmosphere model with the
following parameters: T$_{\rm eff}=6286$~K, a surface gravity of
$\log{g}=3.2$, and a microturbulent velocity of $V_t=3.5$~km s$^{-1}$. 
The model parameters we derived, with the exception of the very low
metallicity we find, are typical of {\it c}/{\it d}-type RR Lyrae's at minimum
light (see also Clementini et al. 1995a). 
	Standard spectroscopic abundance constraints (abundance results
	which show no trends with either line strength or excitation or 
	ionization state), were easily satisfied in the analysis, 
	confirming the adopted photometric temperature.  
Figure~10 shows the
comparison of the spectrum of CU Com near the Mg b lines,
with analogous spectra for two other RR Lyraes (X Ari, [Fe/H]=$-$2.52; and ST
Boo, [Fe/H]=$-$1.80) from Clementini et al. (1995a). Iron-group lines in the 
spectrum of CU Com have strength similar to those in the spectrum of X Ari, 
and are much weaker than those in ST Boo, in agreement with the metallicity 
given by our analysis. 
Note that while weak lines in CU Com appear stronger than in X Ari (due to
the slightly larger metal abundance), the stronger lines (noticeable the Mg b 
ones) 
are somewhat shallower: this is due to the lower microturbulent velocity of CU 
Com (an RRd) with respect to X Ari and ST Boo (both RRab's).
Uncertainties in the derived abundances are mainly due to possible errors in
the atmospheric parameters ($\pm$ 100 K in T$_{\rm eff}$, $\pm$ 0.3 dex in
$\log{g}$, $\pm$ 0.5 km s$^{-1}$ in V$_t$, and $\pm$0.2 dex in [A/H]) and on the
adopted model atmospheres (Kurucz 1993). Our estimate of the random error
contribution (including errors in measuring individual lines) is 0.13 dex.
However, the adopted model atmospheres may contribute an additional 0.15 dex.
Thus, our conservative estimate of the metallicity and of the total error 
for CU Com is [Fe/H]=$-2.38\pm 0.20$. This metallicity is on the same scale of
CG97 and Clementini et al. (1995a) for RR Lyrae variables. On these same
scales X~Ari, the most metal poor RR Lyrae, of any type, known so far, has a
metal abundance of $-$2.52 dex and M15, the most metal poor globular cluster
where RRd variables have been found, has a metal abundance [Fe/H]=$-$2.12. CU
Com thus becomes the most metal deficient double mode RR Lyrae ever detected.

Table~7 shows the abundance ratios with respect to iron we derive for a 
few other elements. As a comparison, in the last column are listed the same 
abundance
ratios obtained by Clementini et al. (1995a) for X Ari: we find a
very similar abundance pattern, namely an overabundance of $\alpha-$elements 
by about 0.4 dex, and a large deficiency of Mn and of Ba. This is the typical
abundance pattern for very metal-poor stars ([Fe/H]$<-2$). 
We also get a very low Al abundance from the resonance lines, 
a result shared by other extremely metal-poor stars (Gratton \& Sneden 1988).
\section{Discussion and conclusions}

CU Com is the sixth double-mode RR Lyrae identified in the field of our Galaxy.
As is true of most of the RRd's known so far, the amplitude of the primary
(first-overtone) period is about twice the amplitude of the secondary
(fundamental) period. 
Final periods and epoch, as well as the average quantities derived for CU Com 
from both the photometric and the spectroscopic analyses 
are 
given in Table~8.
According to the period ratio P$_1$/P$_0$=0.7457  and
the secondary period (P$_0$), CU Com falls on the metal poor side of the
Petersen diagram, in the region populated by the M68 and M15
double-mode pulsators
(see Figure~11).
 This is confirmed by the
extremely low metallicity ([Fe/H]=$-2.38 \pm 0.20$) derived for CU Com 
 from the
abundance analysis of the spectra taken at minimum light, making it the most
metal poor double mode RR Lyrae known so far. 

An estimate for the
mass of CU Com has been obtained from the Petersen diagram, 
following Alcock et al. (1997) and adopting 
Bono et al. (1996) theoretical
mass-calibration. 
Following the suggestion by Cox (1991) that the inclusion of new opacity 
evaluations (Rogers \& Iglesias 1992) in the computation of pulsation models 
allows one to reconcile pulsational and evolutionary predictions for 
the stellar mass of double mode RR Lyrae, Bono et al. (1996) investigated 
the Petersen diagram predicted for Oosterhoff I and Oosterhoff II cluster 
pulsators on the basis of their nonlinear 
convective pulsation models with updated input physics.
The properties of these models, as well as the adopted physical and numerical 
assumptions, are extensively discussed in Bono \& Stellingwerf (1994) and Bono
et al. (1996, 1997a,b).
The main result found by Bono et al. (1996) was that the Petersen diagram 
based on nonlinear computations is able to provide valuable constraints on 
the mass and on the luminosity level of double mode
RR Lyrae belonging to Oosterhoff I and Oosterhoff II clusters. 
  
The comparison between the predictions of Bono et al. (1996) and the location
 of CU Com in the Petersen diagram 
indicates for this star a mass 
larger than 0.80 M$_\odot$ (see Figure~11), placing it among the most 
massive double-mode RR Lyr's. 
In order to better constrain the ``pulsational'' mass of CU Com,
new sequences of nonlinear models with 0.83 and 0.85 M$_\odot$, respectively, 
have been computed, using exactly the same code and the same physical inputs 
as in Bono et al. (1996). The behavior of these new models in the 
Petersen diagram is shown in  Figure~11 (dashed and 
long-dashed lines) for
the computed luminosity levels ($\log {\rm L}/{\rm L}_\odot$ = 1.72 and 1.81
for the 0.83 M$_\odot$ model, and only the $\log {\rm L}/{\rm L}_\odot$ = 1.81 level for 
the 0.85 M$_\odot$ model). CU Com, as well as most of  
the M68 and M15 RRd's, 
is very well fitted by the new pulsational models which indicate   
a mass of 0.830 $\pm$ 0.005 M$\odot$ and a luminosity level
close to 1.81. 
A relation between the
mass and the metallicity is known to exist 
among RRd variables in globular clusters and possibly in the field in our 
Galaxy,
in Local Group dwarf spheroidals,  
and in the LMC (see Bragaglia et al. 2000). CU Com
well fits that relation and extends it to the lowest metallicity end.
In Figure~11 there are a couple of field RRd stars that are less massive then 
those in IC 4499, these are NSV09295 (Garcia-Melendo \& Clement, 1997, 
the lowest filled triangle of Figure 11)
and VIII-10 (Clement et al 1991). No metal abundance is available for 
NSV09295, while according to its metallicity VIII-10 is about 0.3 dex more
metal poor than IC 4499 (see Introduction). Indeed, VIII-10 falls slightly off 
the mass-metallicity relation in Figure 7 of Bragaglia et al 2000. However, 
given the small number of RRd's in the field of our Galaxy 
we think that no firm conclusion can be reached on whether 
the Galactic field RRd's do or do not actually follow the same
 mass-metallicity relation of the cluster RRd's. 

According to the evolutionary models consistent with 
ours (Cassisi et al. 1998, 1999) a stellar mass of 
about 0.8 M$_\odot$ is expected to populate the RR Lyrae gap for a 
metallicity Z=0.0001.
Hence, the expected stellar mass populating 
the RR Lyrae instability strip in metal-poor globular clusters such as
M15 and M68 is close to 0.80 M$\odot$ with a luminosity level 
between 1.75 and 1.8, depending on the
assumed efficiency of element diffusion.
Model isochrones by the same authors (Cassisi et al. 1998, 1999) suggest a 
turn-off age of 
11.6 Gyr and a turn-off mass of 0.80 M$\odot$ for a red giant mass   of 
0.83~M$_{\odot}$ (when diffusion and mass loss are neglected) with the typical 
low metallicities of 
clusters such as M15. 

Taking into account an $\alpha$-enhancement contribution of 
$\sim$ 0.4 dex, which is typical of the lowest metallicity stars 
in the 
solar neighborhood  and in globular clusters such as, for instance, M68 (see 
e.g. Carney 1996) which is 
also shared by 
CU Com (see Section 4.2), 
and including it in the global metallicity computation following 
the relation provided by Salaris et al. (1993),
one would obtain for these stars a metallicity Z=1.5 $\times 10^{-4}$.
This tiny metallicity increase with respect to Z=0.0001 is not expected to
change the results of the pulsation analysis. On the other hand, model
isochrones for Z=0.0002 predict an age ranging from 10.9 to 11.2 Gyr,
depending on the efficiency of element diffusion,  and a turn-off mass of 
0.80~M$\odot$
 for a red giant mass   of 
0.83~M$\odot$. 
Indeed, adopting this metallicity Cassisi et al. (1999)  estimated an age 
of 11 $\pm$ 1 Gyr
for M68.      
It is also worth noticing that 
the luminosity level of CU Com, of the M68 and M15 RRd's inferred from 
the Petersen approach is in 
good agreement with the evolutionary predictions
one may extrapolate, for a mass of $\sim$  0.83 M$_\odot$,
 from updated horizontal branch evolutionary models (Cassisi et al. 1998).

Thus, if the pulsational masses estimated from 
Figure~11
 are correct, the M68 and M15 RRd's 
should have masses around 0.83 M$\odot$ and luminosities of about 
$\log {\rm L}/{\rm L}_\odot$= 1.81.
This would imply that almost no mass loss
occurs during the red giant branch evolution of these very metal poor cluster 
stars.

Unfortunately, no firm constraint on the red giant phase mass loss can be 
inferred from the large pulsational mass of CU Com because, at variance with 
the
double mode pulsators in globular clusters, we do not have any information
on its age, hence, on its turn-off mass.
However, (i) given the similarity of CU Com mass with masses of double 
mode pulsators in M68 and M15, (ii) given its extremely low metal abundance,
 and (iii) given the direct dependence of mass loss upon metal abundance 
predicted by canonical mass loss relations such as those by Reimers (1975)
or Maeder (1992), it is probable that CU Com has not lost 
a large amount of mass during its red giant branch evolution.

The synthetic horizontal branch distributions computed by Rood (1973) 
first showed the need 
for an average mass loss of about 
0.2 M$_\odot$ prior to the horizontal branch phase, coupled to a $\sim$ 10$\%$
dispersion of 0.025 M$_\odot$ in order to reproduce the color extension
of the horizontal branch of metal poor globular clusters (see also
Renzini \& Fusi Pecci 1988, Chiosi 1998). 
Reimers (1975) empirical formula 
predicts that the mass loss needed to reproduce the observed horizontal branch 
morphologies at Z $\simeq$ 0.001 is obtained for a value  of 0.40 $\pm$ 0.04 
for the mass 
loss  rate efficiency parameter $\eta$ (see Renzini \& Fusi Pecci 1988).
Thus, if the 0.83 M$_\odot$ we derive for CU Com is correct, its turn-off mass
could range from 0.83 M$_\odot$, in the case of no mass loss, up to 
0.93 M$_\odot$ for an assumed mass loss of 0.1 M$_\odot$ (a reasonable 
assumption given the overall uncertainties). Correspondingly, CU Com age could
range  from $\sim$ 10 to 6.5 Gyr: the turn off
ages of a 0.83 
and a 0.93 M$_\odot$, respectively, according to Cassisi et al. (1998) models 
at  Z=0.0002 and including diffusion.
 
In this framework,
the implication of the high mass and low metallicity
of CU Com on the evolutionary interpretation of the double-mode phenomenon 
is that, if  
its estimated 
metallicity and mass are correct, CU Com is a massive,  
overluminous RR Lyrae almost as old as M68, sharing the typical 
$\alpha$-elements overabundance of very metal-poor stars,   
which started its horizontal branch 
evolution on the very red side of the instability strip and is now moving 
blueward 
across the strip while evolving from the {\it ab} to the {\it c} type
region.

ACKNOWLEDGEMENTS

We warmly thank P. Montegriffo for his expert advice during 
the use of the 
program GRATIS and for his readiness to modify parts of the software to better 
meet specific needs during the analysis of CU Com.
This work was partially supported by MURST-Cofin98 under the project
"Stellar Evolution". 
S.DiT. was partially supported with an Italian CNAA fellowship 
at the University of Texas at Austin.
H.S. and C.S. thank the United States National Science Foundation for support 
under grants AST-9528080 and AST-9618364.

\newpage

\figcaption[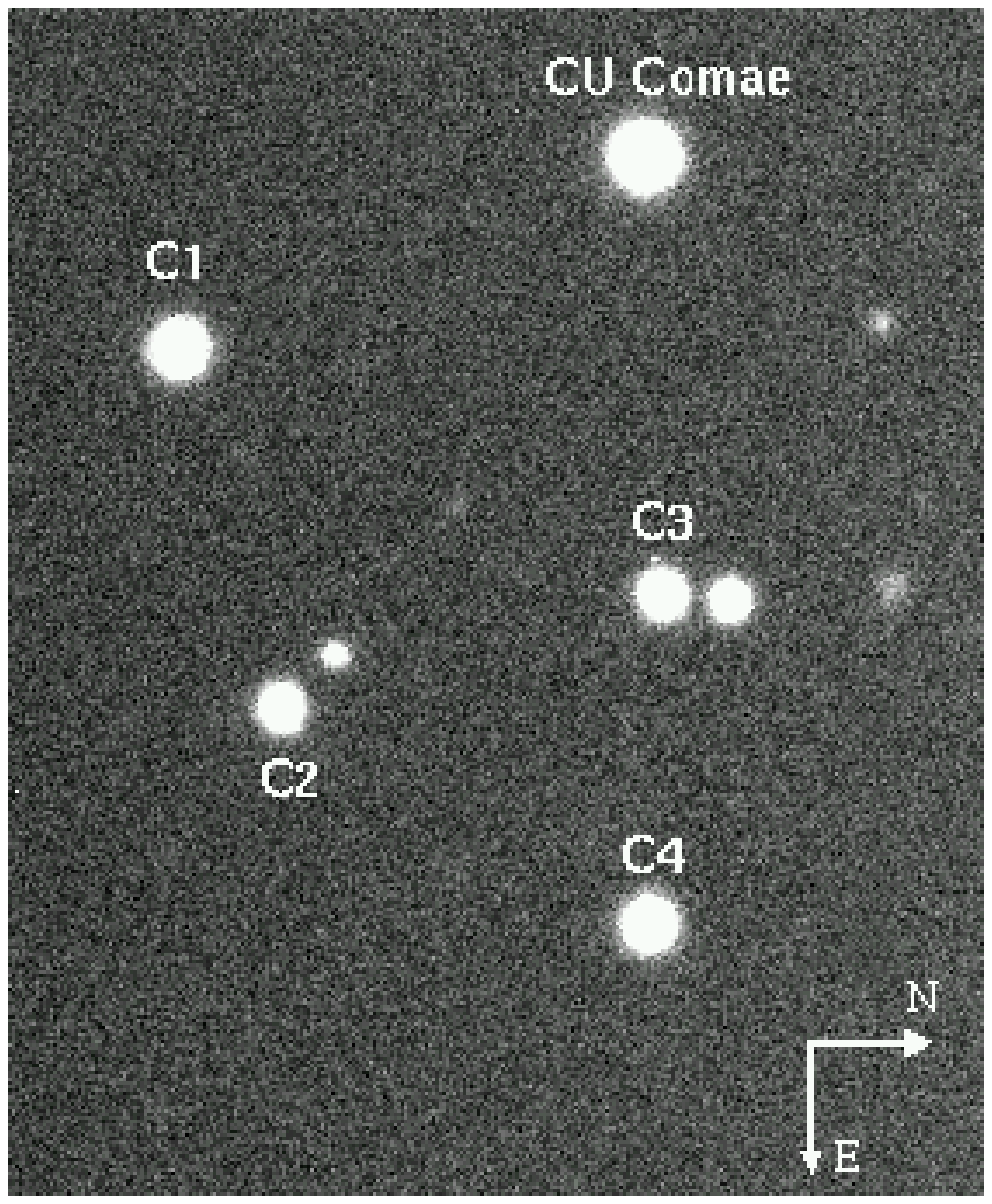]{A 2.3 $\times$ 2.8 
arcmin$^2$ CCD image of
the CU Com field.  
The four stars marked beside CU Com are non 
variable objects which were used as reference stars to obtain 
differential measurements of the magnitude variation of CU Com.}

\figcaption[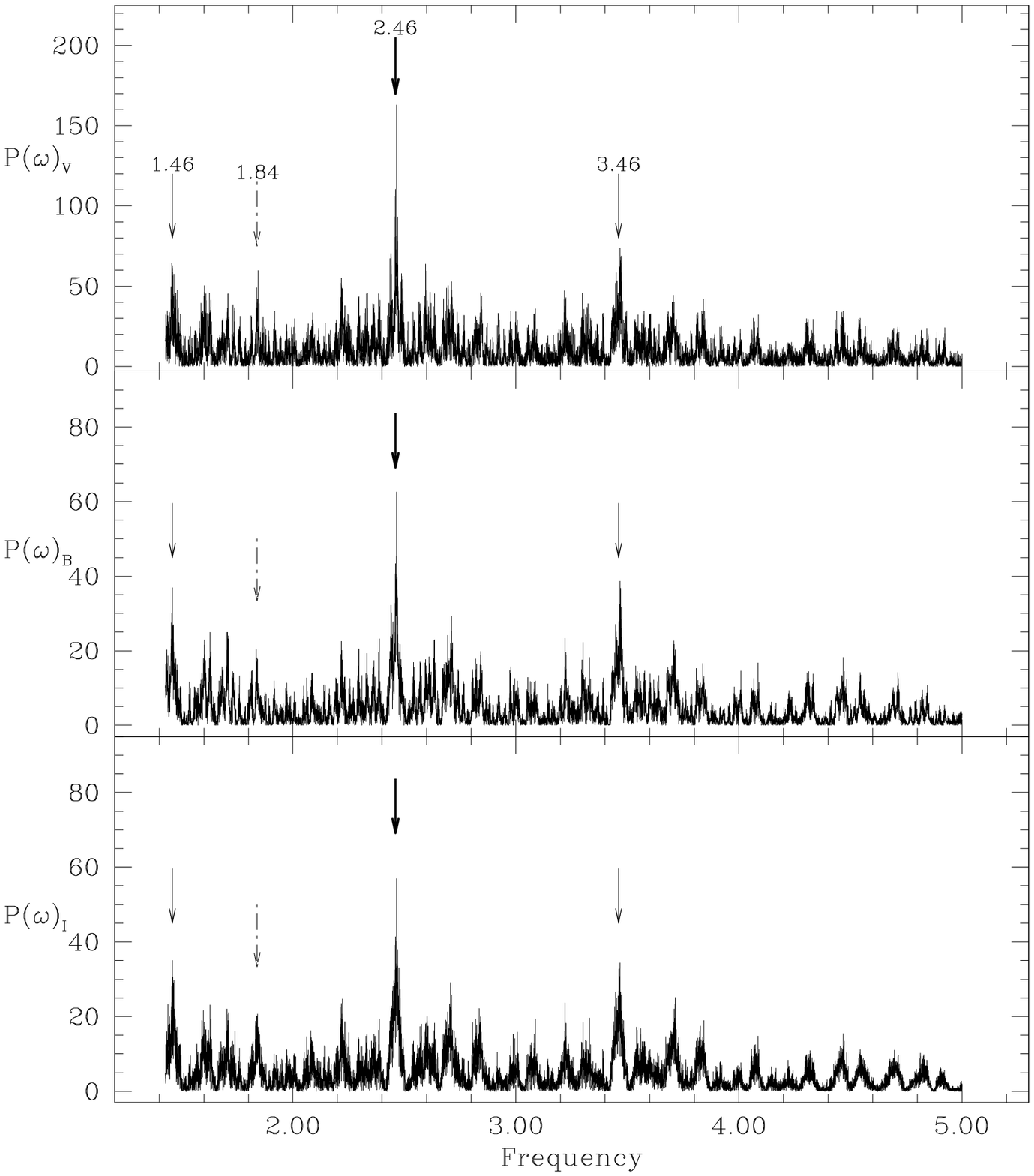]{Periodograms of the V, B, and I  data of CU Com which 
identify the
most probable frequency of the data ($\omega$=2.46), the two aliases at
$\omega$=1.46 and $\omega$=3.46, and a secondary periodicity
at $\omega$=1.84 (see text).}

\figcaption[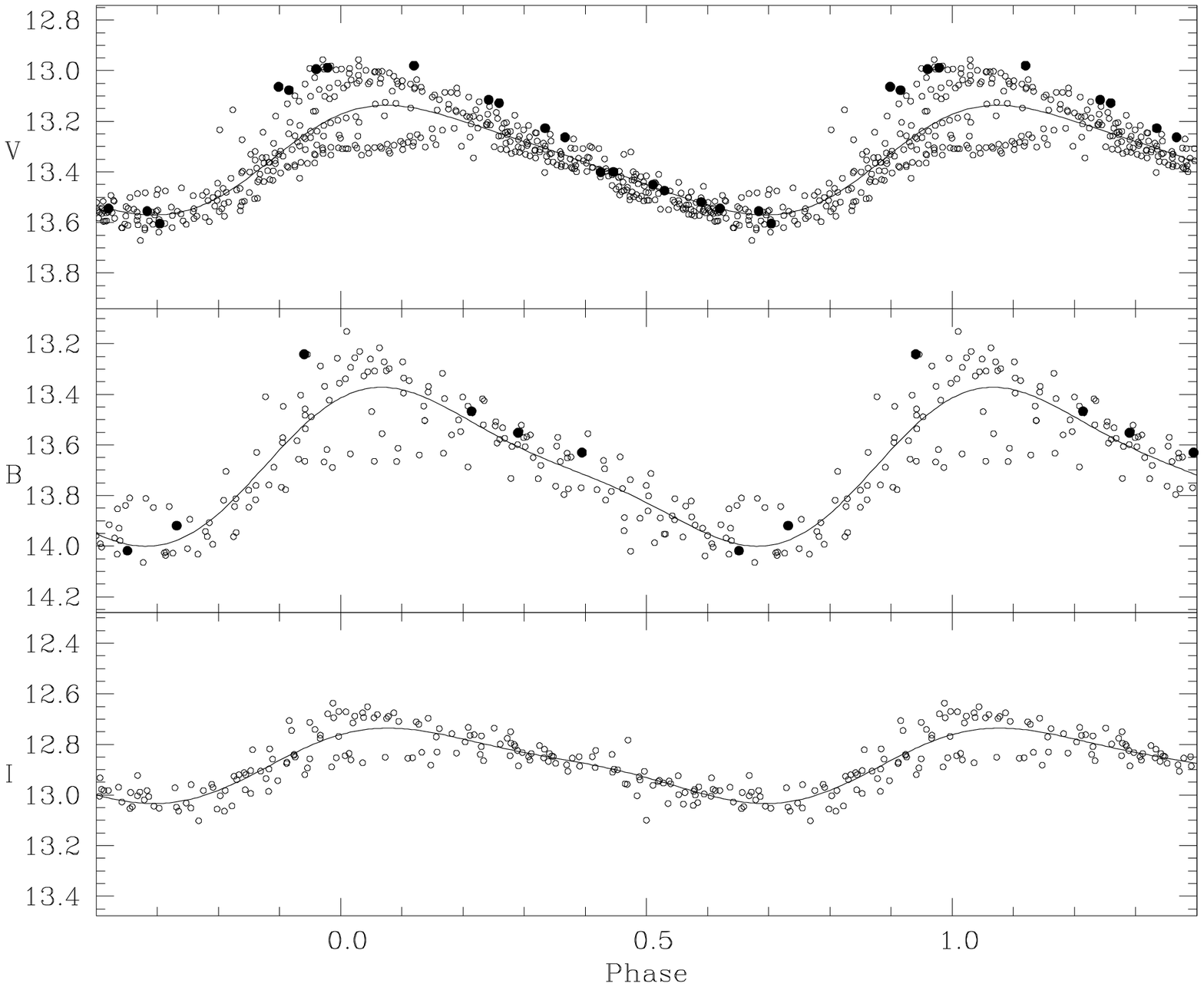]{V, B, and I light curves of CU Com. Data are phased according to a primary
(first-overtone) period of pulsation P$_1$=0.$^d$405759. The complete data-set
spanning 11 years of observations (1989-1999) is plotted. Solid lines are the
best fitting models used to fit the data. Filled circles mark photometric
observations obtained with the 40 cm telescope simultaneous to the spectroscopic 
ones.}

\figcaption[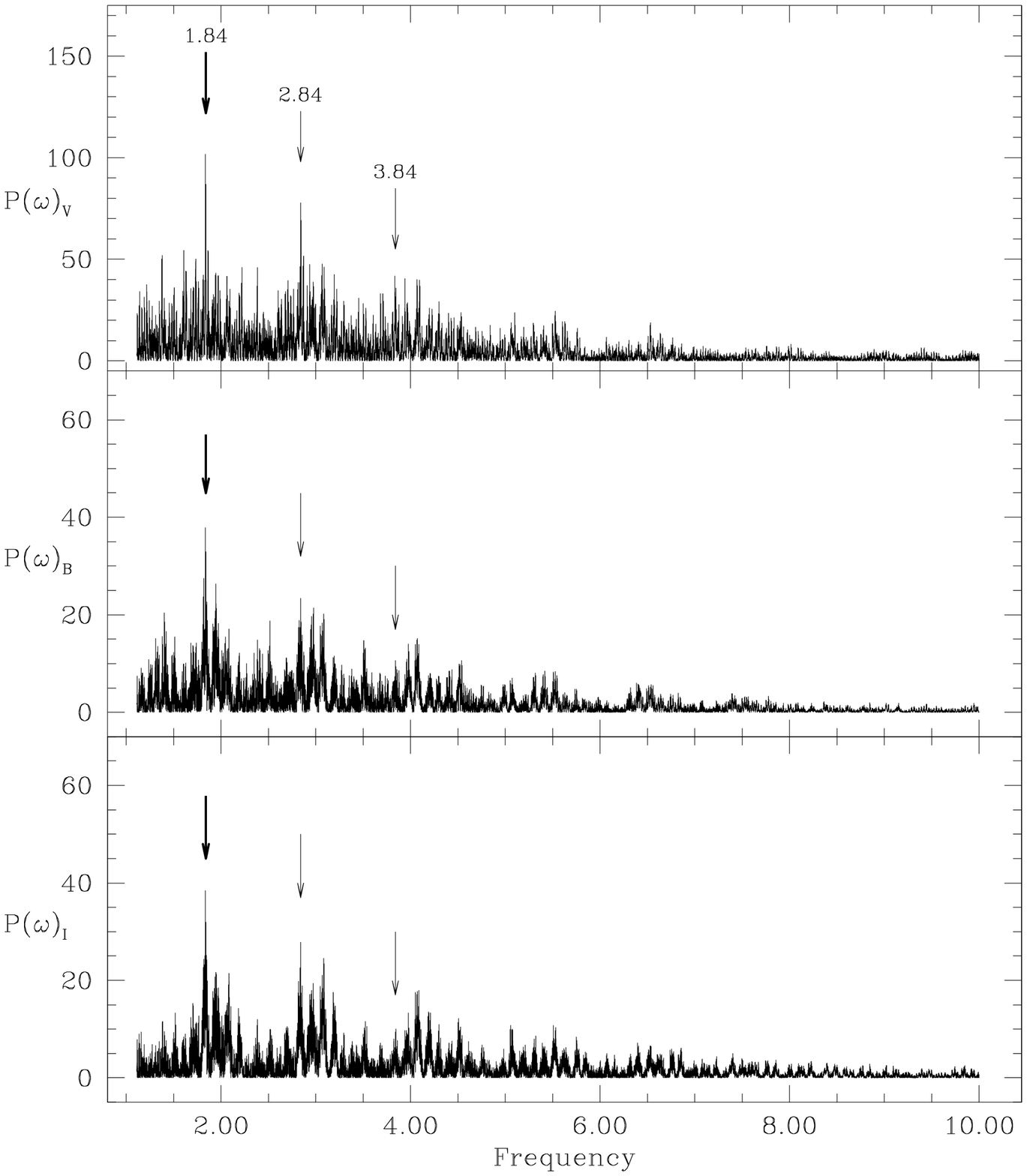]{Periodograms of the residuals of the V, B and I data, respectively, 
which identify the most probable secondary frequency of the data 
($\omega$=1.84) and the two aliases at $\omega$=2.84 and
$\omega$=3.84 (see text).}

\figcaption[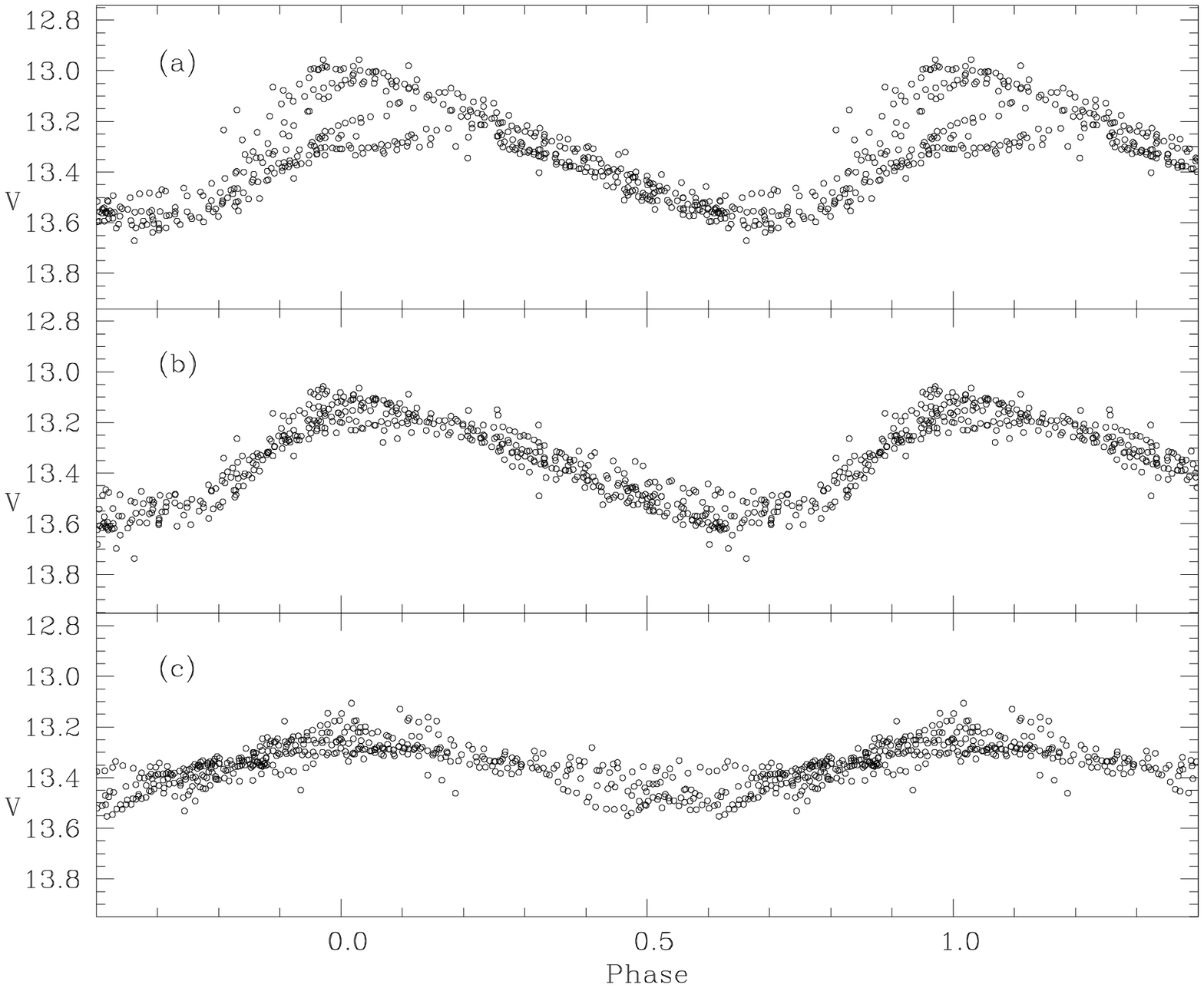]{Light curve of the V data of CU Com phased according to the
primary (first-overtone) period of pulsation P$_1$=0.$^d$4057605 derived at the end of
the prewhitening procedure (top panel). The central panel shows the light
curve of the primary (first-overtone) period after prewhitening of the
secondary (fundamental) period P$_0$=0.$^d$5441641, and the bottom panel shows the
light curve of the secondary (fundamental) period after prewhitening of the
primary (first-overtone) period.}

\figcaption[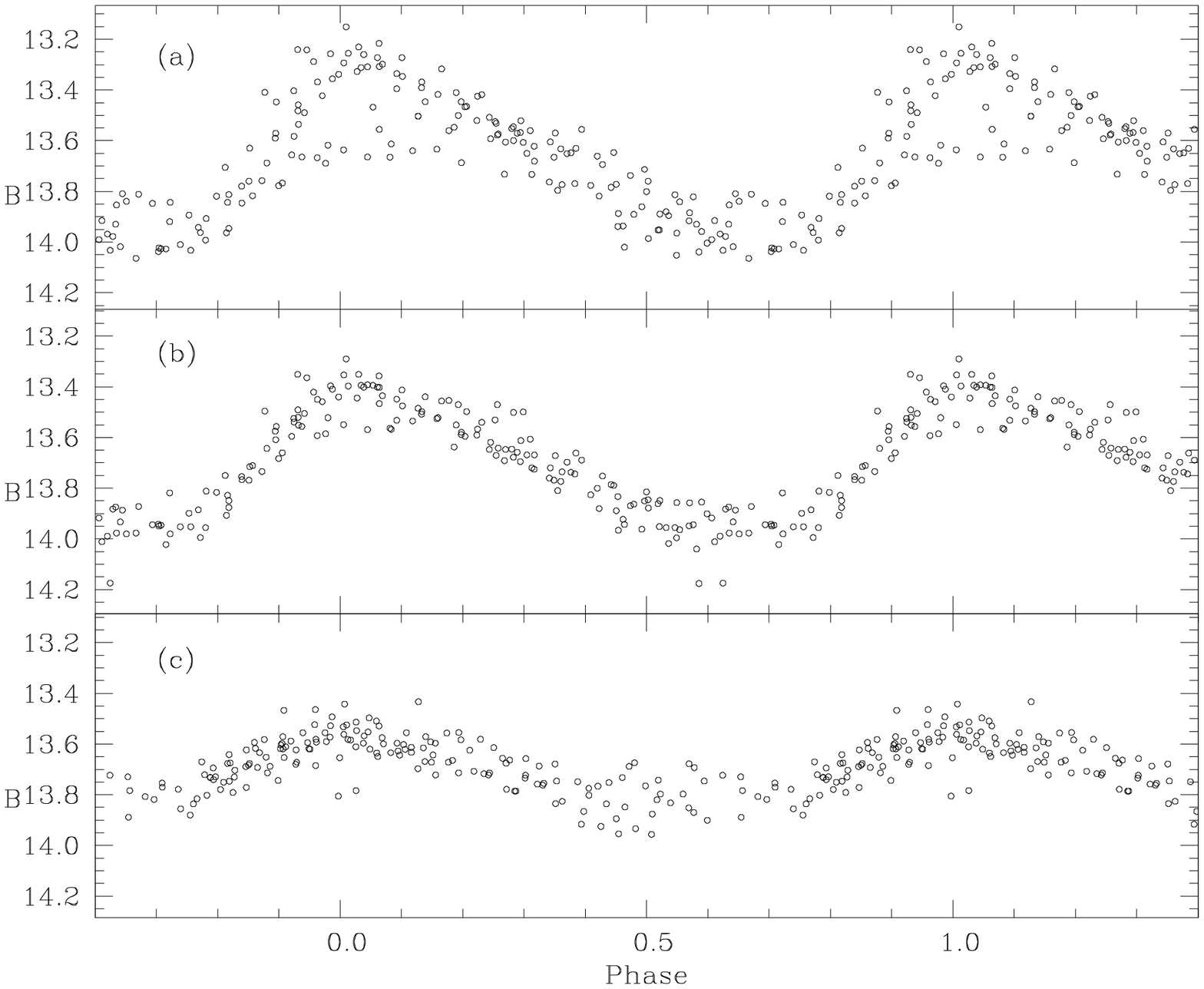]{Same as Figure 5 for the B data-set of CU Com.}

\figcaption[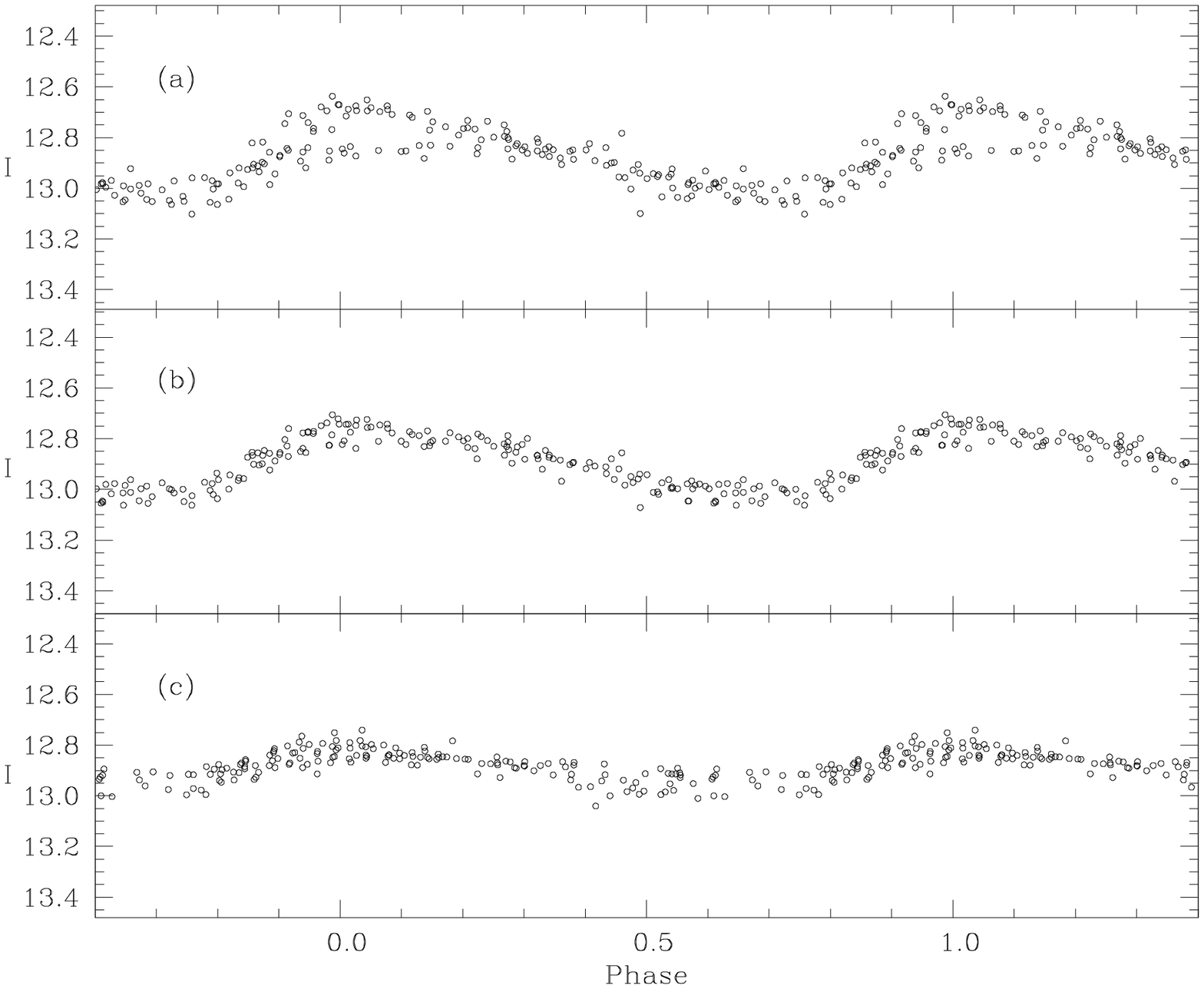]{Same as Figure 5 for the I data-set of CU Com.}

\figcaption[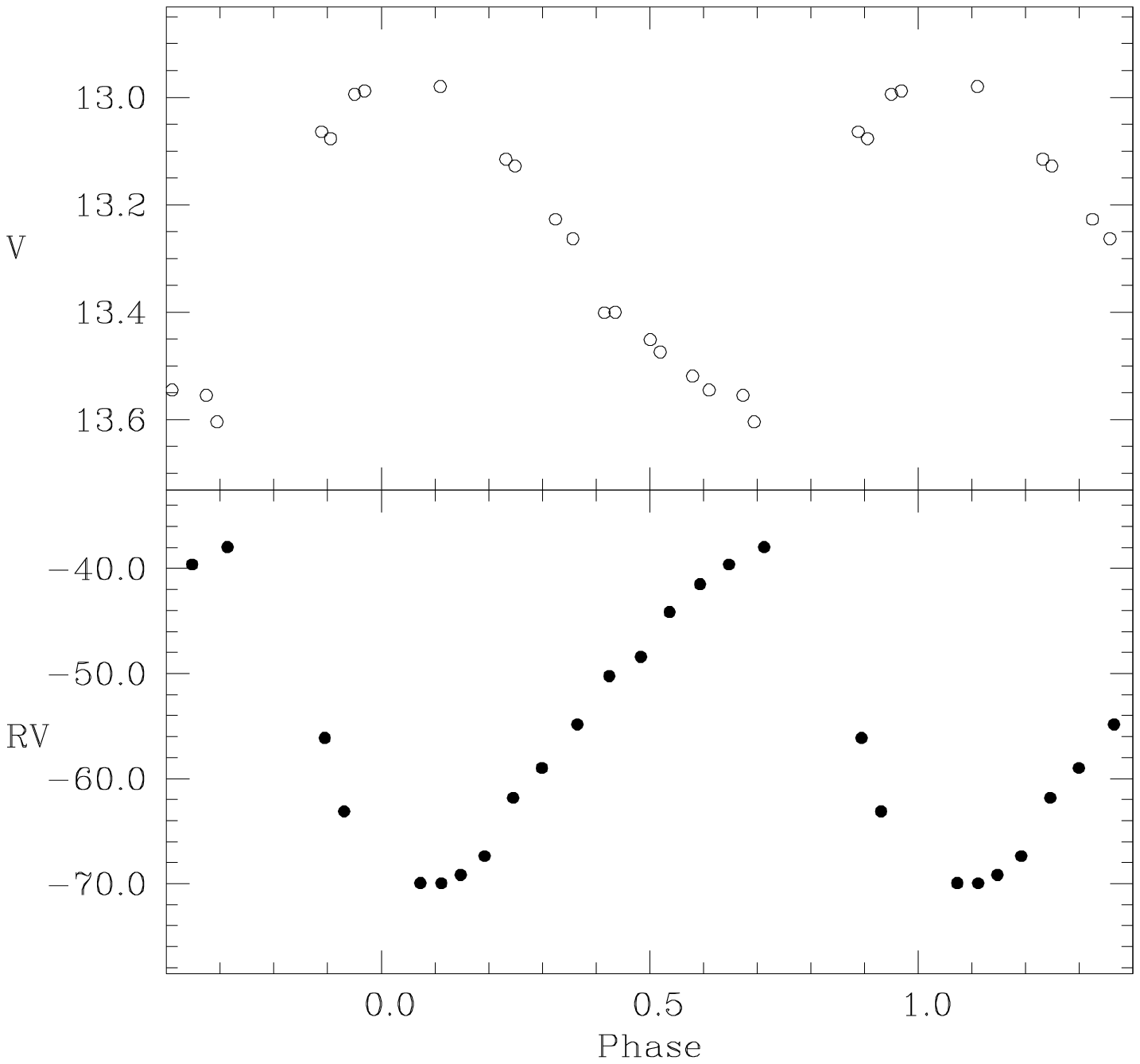]{Radial velocity curve (bottom panel) and simultaneous 
V light curve of CU Com (top panel), obtained with the 2.7 m McDonald telescope and 
the 40 cm of the Southwestern University, respectively.}

\figcaption[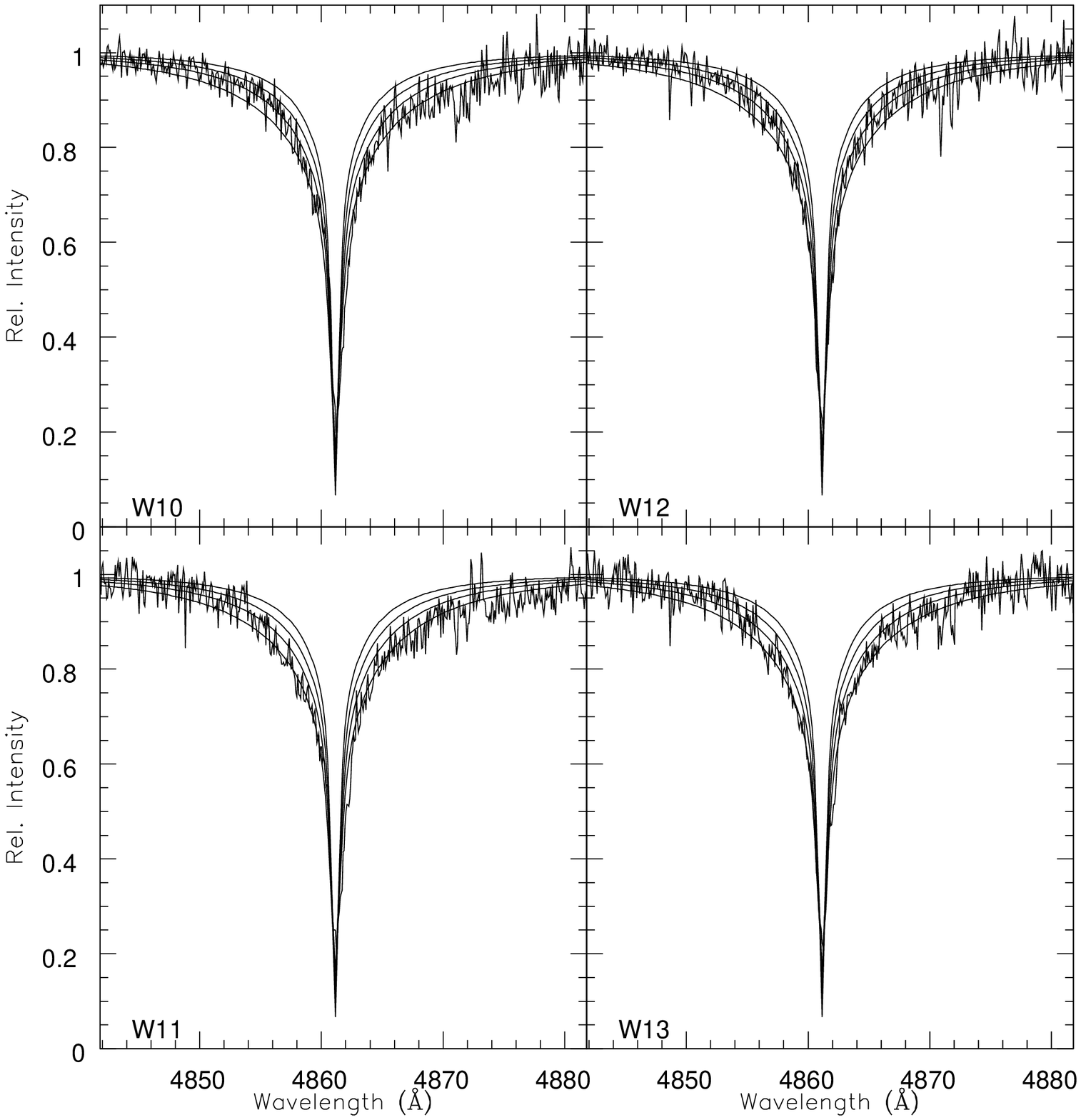]{Comparison of the observed H$\beta$\ line profiles
and synthetic spectra computed using the Kurucz (1993) model atmospheres, with the
overshooting option switched on (see text). Labels indicate the 
spectra according to their identification in Table 5. The average
temperatures we obtained for the four spectra taken close to minimum light is
6400~K$\pm 150$ K.}

\figcaption[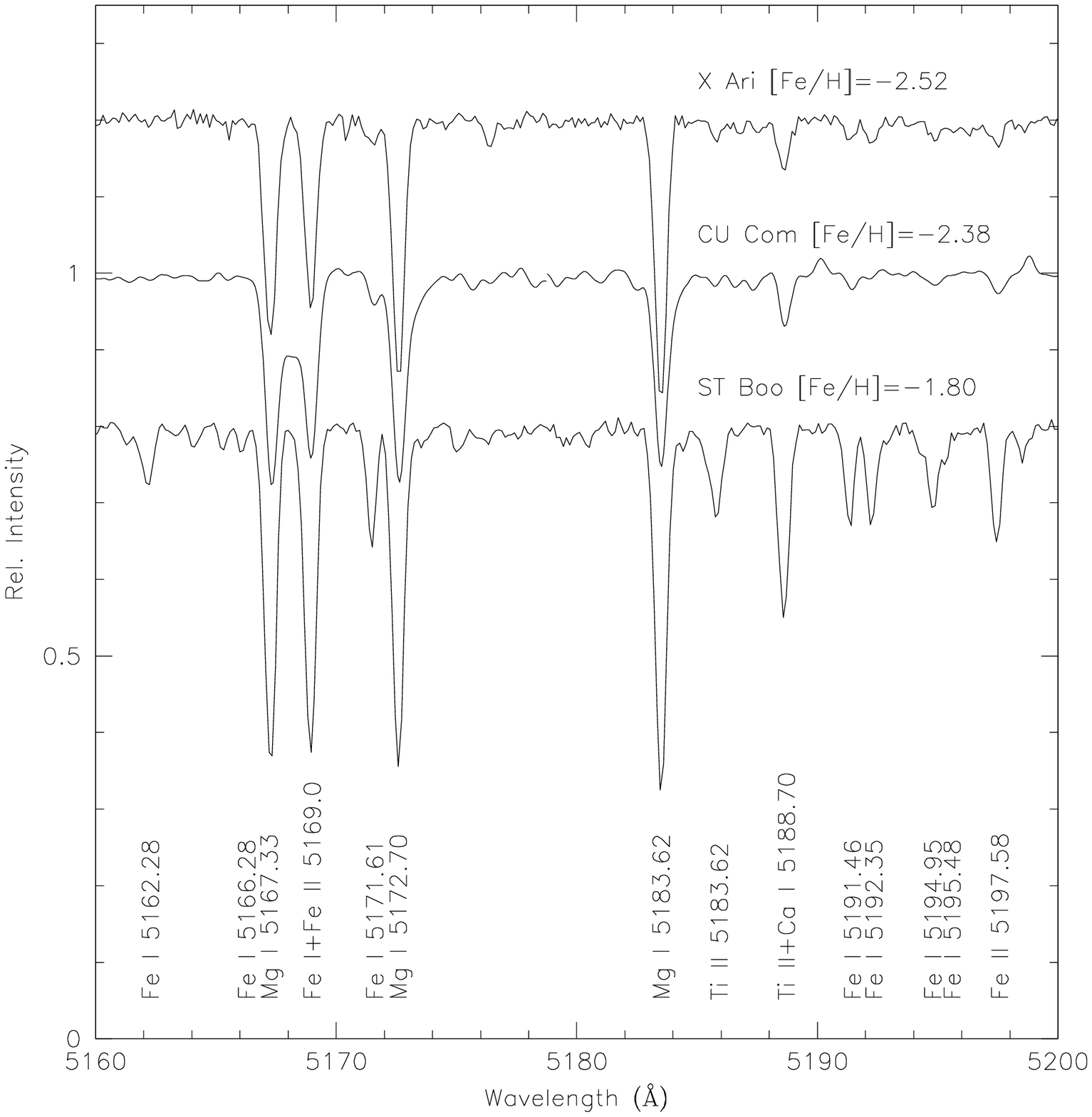]{A small section of the coadded spectrum of CU Com (near the Mg b lines),
compared with spectra for two other RR Lyraes (X Ari, [Fe/H]=$-$2.52; 
and ST Boo, [Fe/H]=$-$1.80) from Clementini et al. (1995a). These two latter  
have been shifted vertically for clarity. Note that the iron-group lines in
the spectrum of CU Com have strength similar to those in the spectrum of X Ari, 
and are much weaker than those in ST Boo, in agreement with the metallicity 
given by our analysis.}

\figcaption[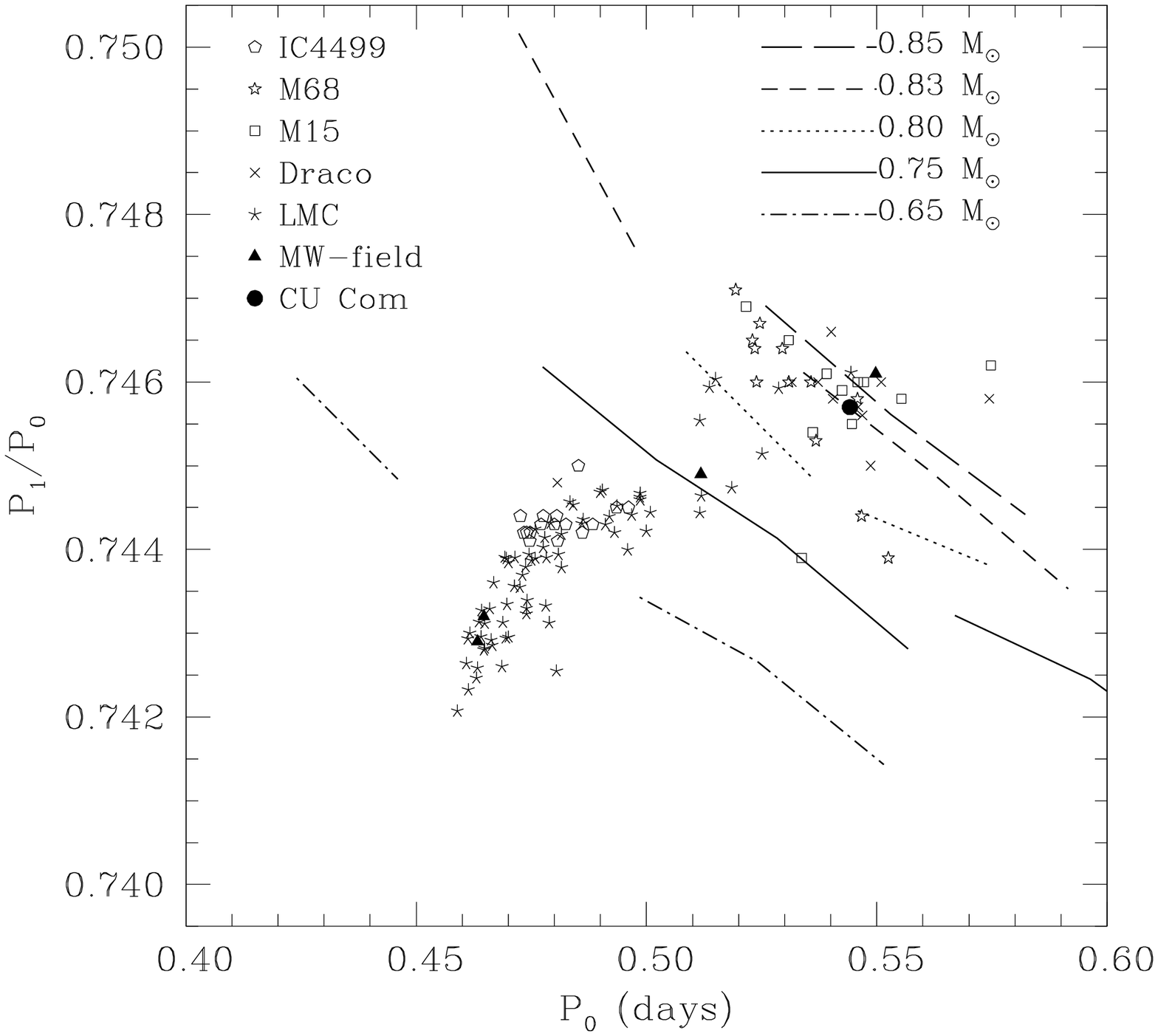]{Position of CU Com on the Petersen diagram. Data shown in 
the figure are from  Walker \& Nemec (1996; IC4499), 
Walker (1994; M68), Nemec (1985b; M15), Nemec (1985a; Draco),
Clement, Ferance \& Simon (1993; AQ Leo, VIII-10, and VIII-58), 
Garcia-Melendo \& Clement (1997; NSV09295), and Alcock et al. (1997; LMC).
Dashed-dotted, solid and dotted lines represent Bono et al. (1996) pulsational 
models for masses of 0.65, 0.70 and 0.80 M$\odot$, and 
$\log$ L/L$\odot$= 1.72 (upper portion of the curves) and $\log$ 
L/L$\odot$= 1.81 (lower portions), respectively. 
The dashed and long-dashed 
lines show our new pulsational models  
for 0.83 M$\odot$ and 0.85 M$\odot$, respectively. Only the 
$\log {\rm L}/{\rm L}_\odot$= 1.81 luminosity level is shown in the 
latter case.}

\newpage
\begin{table}
\begin{center}
\caption{Journal of the new photometric observations}
\begin{tabular}{c c c c c}
\tableline
\tableline
\multicolumn{1}{c}{Year} &
\multicolumn{3}{c}{N.of Observations}&
\multicolumn{1}{c}{Observed intervals}\\
     &   B  &  V  &  I & (HJD$-$2452400000)   \\
\tableline
1995 &   8  &  16 &   7  & 49789 - 49846 \\
1996 &  23  &  21 &  22  & 50130 - 50195  \\
1997 &  14  &  27 &  16  & 50550 - 50551 \\
1998 &   1  &   1 &      & 50842 \\
1999 & 126  & 281 & 113  & 51220 - 51301 \\
Tot  & 172  & 346 & 158  & \\
\tableline
\end{tabular}
\end{center}
\label{t:tab1}
\end{table}

\newpage
\begin{table}
\begin{center}
\caption{Magnitudes of the comparison stars}
\begin{tabular}{c c c c c}
\tableline
\tableline
\multicolumn{1}{c}{Star} &
\multicolumn{1}{c}{N$_{GSC}$}&
\multicolumn{1}{c}{V} &
\multicolumn{1}{c}{B} &
\multicolumn{1}{c}{I} \\
\tableline
C1 & 0144701247 &14.08 $\pm$ 0.03 & 14.89 $\pm$ 0.03 & 13.19 $\pm$ 0.04 \\
C2 & 0144700851 &15.27 $\pm$ 0.03 & 15.88 $\pm$ 0.03 & 14.62 $\pm$ 0.04 \\
C3 & 0144701207 &15.12 $\pm$ 0.03 & 15.80 $\pm$ 0.03 & 14.40 $\pm$ 0.04 \\
C4 & 0144701193 &14.60 $\pm$ 0.03 & 15.31 $\pm$ 0.03 & 13.87 $\pm$ 0.04 \\
\tableline
\end{tabular}
\vskip 0.5 cm
\end{center}
\label{t:tab2}
\end{table}

\newpage
\begin{table}
\begin{center}
\caption{The photometric observations.\tablenotemark{*}}
\begin{tabular}{r c r c r c}
\tableline
\tableline
\multicolumn{1}{c}{~~HJD}&
\multicolumn{1}{c}{~$\Delta{\rm V}$}&
\multicolumn{1}{c}{~~HJD}&
\multicolumn{1}{c}{~$\Delta{\rm B}$}&
\multicolumn{1}{c}{~~HJD}&
\multicolumn{1}{c}{~$\Delta{\rm I}$}\\
($-$2400000) & & ($-$2400000) & & ($-$2400000) &\\
\tableline
49789.3640137 &~~$-$0.676~~&49789.3725840   &~~$-$1.127~~& 49789.3876953
&~~$-$0.242~~\\
    0.3811035 &~~$-$0.665~~&      0.4024110 &~~$-$1.065~~&     0.4169922
    &~~$-$0.209~~\\
    0.3942871 &~~$-$0.633~~&      0.4328630 &~~$-$1.048~~&     0.4472656
    &~~$-$0.204~~\\
    0.4106445 &~~$-$0.618~~&      0.4618910 &~~$-$1.044~~&     0.4763184
    &~~$-$0.227~~\\
    0.4233398 &~~$-$0.582~~& & & & \\
    0.4409180 &~~$-$0.589~~& & & & \\
    0.4536133 &~~$-$0.585~~& & & & \\
    0.4699707 &~~$-$0.612~~& & & & \\
\tableline
\end{tabular}
\end{center}
\label{t:tab3}
\tablenotetext{*}{The complete version of this table which includes 
also the updated version of Clementini et al (1995) photometry is in the 
electronic edition of 
the Journal. The printed edition contains only a sample of the present new 
photometry}
\end{table}

\newpage

\begin{table}
\begin{center}
\caption{The heliocentric radial velocities}
\begin{tabular}{l r c c c}
\tableline
\tableline
\multicolumn{1}{c}{Spectrum} &
\multicolumn{1}{c}{~~HJD} &
\multicolumn{1}{c}{$\Phi$~~~} &
\multicolumn{1}{c}{${\rm RV}$~} &
\multicolumn{1}{c}{Error}\\
\multicolumn{1}{c}{} &
\multicolumn{1}{c}{} &
\multicolumn{1}{c}{} &
\multicolumn{1}{c}{km s$^{-1}$} &
\multicolumn{1}{c}{km s$^{-1}$}\\
\tableline
~~~W1& 51222.750103 & 0.073& $-$69.94 & 3.97\\
~~~W2&    0.765833 & 0.112 & $-$69.96 & 2.94\\
~~~W3&    0.780487 & 0.148 &$-$69.17 & 3.75\\
~~~W4&    0.798555 & 0.192 &$-$67.37 & 3.54\\
~~~W5&    0.820304 & 0.246 &$-$61.83  & 2.90\\
~~~W6&    0.841913 & 0.299 &$-$58.99  & 1.84\\
~~~W7&    0.868628 & 0.365 &$-$54.85  & 2.94\\
~~~W8&    0.892935 & 0.425 &$-$50.24  & 3.77\\
~~~W9&    0.916663 & 0.483 &$-$48.41  & 4.45\\
~~~W10&    0.938504 & 0.537 &$-$44.14  & 3.71\\
~~~W11&    0.961503 & 0.594 &$-$41.50  & 2.14\\
~~~W12&    0.983263 & 0.647 &$-$39.62  & 1.99\\
~~~W13&51223.009931 & 0.713 &$-$37.96  & 2.64\\
~~~W14&    0.894986 & 0.895 &$-$56.13  & 2.56\\
~~~W15&    0.909752 & 0.931 &$-$63.12  & 4.68\\
\tableline
\end{tabular}
\end{center}
\label{t:tab4}
\end{table}

\newpage

\begin{table}
\begin{center}
\caption{Temperatures derived from CU Com dereddened colors at minimum light}
\begin{tabular}{c c c c c c}
\tableline
\tableline
\multicolumn{1}{c}{Spectrum} &
\multicolumn{1}{c}{$\Phi$}&
\multicolumn{1}{c}{(B$-$V)$_0$} &
\multicolumn{1}{c}{(V$-$I)$_0$}&
\multicolumn{1}{c}{T$_{\rm eff}$}&
\multicolumn{1}{c}{T$_{\rm eff}$}\\
\multicolumn{1}{c}{} &
\multicolumn{1}{c}{}&
\multicolumn{1}{c}{} &
\multicolumn{1}{c}{}&
\multicolumn{1}{c}{(B$-$V)$_0$}&
\multicolumn{1}{c}{(V$-$I)$_0$}\\
\tableline
W10 & 0.537 & 0.369 & 0.495 & 6305 & 6550\\
W11 & 0.594 & 0.400 & 0.506 & 6130 & 6513\\
W12 & 0.647 & 0.411 & 0.525 & 6070 & 6448\\
W13 & 0.713 & 0.355 & 0.519 & 6385 & 6470\\
\tableline
\end{tabular}
\vskip 0.5 cm
\end{center}
\label{t:tab5}
\end{table}

\newpage
\begin{table}
\begin{center}
\caption{Linelist and measured EW's}
\begin{tabular}{c c c r r c}
\tableline
\tableline
\multicolumn{1}{l}{Element}&
\multicolumn{1}{c}{$\lambda$}&
\multicolumn{1}{c}{E.P.}&
\multicolumn{1}{r}{$\log gf$}&
\multicolumn{1}{c}{EW}&
\multicolumn{1}{c}{$\log n$}\\
\tableline
Fe I  &  4005.24 &  1.56 &  --0.57~ &  111.10 &  5.28\\
Fe I  &  4045.81 &  1.49 &   0.22~ &  159.10 &  5.23\\
Fe I  &  4071.74 &  1.61 &  --0.02~ &  137.50 &  5.23\\
Fe I  &  4132.06 &  1.61 &  --0.63~ &   86.10 &  5.03\\
Fe I  &  4143.87 &  1.56 &  --0.44~ &  102.70 &  5.03\\
Fe I  &  4235.95 &  2.43 &  --0.34~ &   47.90 &  5.06\\
Fe I  &  4383.56 &  1.49 &   0.20~ &  159.50 &  5.13\\
Fe I  &  4415.13 &  1.61 &  --0.61~ &  109.40 &  5.28\\
Fe I  &  4427.32 &  0.05 &  --3.04~ &   26.40 &  5.23\\
Fe I  &  4459.14 &  2.18 &  --1.28~ &   14.80 &  5.13\\
\tableline
\end{tabular}
\end{center}
\end{table}

\newpage
\begin{table}
\caption{Average elemental abundances for CU Com and X Ari}
\begin{tabular}{lcrcr}
\tableline
\tableline
               &    &CU Com &      & X Ari \\
               & n & mean~~  &r.m.s &       \\
\tableline
$[$Fe/H$]$ I   & 15 & $-$2.35~~ & 0.10 & $-$2.52~ \\
$[$Fe/H$]$ II  &  8 & $-$2.40~~ & 0.08 & $-$2.48~ \\
\\
$[$Mg/Fe$]$ I  &  5 &    0.39~~ & 0.29 &    0.52~ \\
$[$Al/Fe$]$ I  &  2 & $-$0.99~~ & 0.07 & $-$0.30~ \\
$[$Ca/Fe$]$ I  &  7 &    0.29~~ & 0.15 &    0.45~ \\ 
$[$Sc/Fe$]$ II &  4 &    0.24~~ & 0.21 &    0.31~ \\
$[$Ti/Fe$]$ II & 11 &    0.46~~ & 0.15 &    0.34~ \\
$[$Cr/Fe$]$ I  &  1 & $-$0.22~~ &      & $-$0.30~ \\
$[$Mn/Fe$]$ I  &  3 & $-$0.83~~ & 0.14 & $-$0.81~ \\
$[$Sr/Fe$]$ II &  2 & $-$0.21~~ & 0.04 &         \\
$[$Ba/Fe$]$ II &  1 & $-$0.71~~ &      & $-$0.75~ \\
\tableline
\end{tabular}
\label{t:elements}
\end{table}

\newpage
\begin{table}
\begin{center}
\caption{Properties of CU Com}
\begin{tabular}{c c}
\tableline
\tableline
Type  & RRd \\
${\rm [Fe/H]}$& $-2.38 \pm 0.20$\\
Epoch & 2450142.$^d$5860390 \\
P$_1$ & 0.$^d$4057605 $\pm$ 0.0000018 \\
P$_0$ & 0.$^d$5441641 $\pm$ 0.0000049 \\
P$_1$/P$_0$ &0.745658 $\pm$ 0.000007 \\
M & 0.830$\pm0.005$M$\odot$\\
$< V >$ & 13.34\\
$< B >$ & 13.67\\
$< I >$ & 12.88\\
$< B > - < V >$ & 0.33\\
$< V > - < I >$ & 0.46\\
A$_1$(V) & 0.43\\
A$_0$(V) & 0.22\\
A$_1$(B) & 0.55\\
A$_0$(B) & 0.25\\
A$_1$(I) & 0.27\\
A$_0$(I) & 0.14\\
A$_{\rm RV}$ & ~~32.73 km s$^{-1}$\\
$\gamma$ & $-$54.13 km s$^{-1}$\\
\tableline
\end{tabular}
\end{center}
\label{t:tab6}
\end{table}

\end{document}